# Phase Steps and Hot Resonator Detuning in Microresonator Frequency Combs

Pascal Del'Haye[1*], Aurélien Coillet[1], William Loh[1], Katja Beha[1], Scott B. Papp[1], and Scott A. Diddams[1]

[1]National Institute of Standards and Technology (NIST), Boulder, CO 80305, USA

**Experiments and theoretical modeling yielded significant progress towards understanding of Kerr-effect induced optical frequency comb generation in microresonators. However, the simultaneous interaction of hundreds or thousands of optical comb frequencies with the same number of resonator modes leads to complicated nonlinear dynamics that are far from fully understood. An important prerequisite for modeling the comb formation process is the knowledge of phase and amplitude of the comb modes as well as the detuning from their respective microresonator modes. Here, we present comprehensive measurements that fully characterize optical microcomb states. We introduce a way of measuring resonator dispersion and detuning of comb modes in a hot resonator *while* generating an optical frequency comb. The presented phase measurements show unpredicted comb states with discrete $\pi$ and $\pi/2$ steps in the comb phases that are not observed in conventional optical frequency combs.**

Optical frequency combs have proven to be powerful metrology tools for a variety of applications as well as for basic research [1, 2]. Conventional optical frequency combs are based on mode-locked lasers, for which a theoretical framework of the underlying comb generation mechanism has been well established [3, 4]. A miniaturized platform for optical frequency comb generation in optical microresonators has been shown in the past years [5-16] and would be a valuable component of future on-chip optical circuits [17]. The comb generation process itself can be described as a cascade of four-wave mixing processes. In the first step, light from a continuous wave pump laser is converted into equidistant sidebands by degenerate four-wave mixing obeying $2f_p = f_s + f_i$ (with pump laser frequency $f_p$ and sideband frequencies $f_{s,i}$). At a later stage in the comb generation process, comb modes start to interact and generate new lines via non-degenerate four-wave mixing obeying energy conservation with $f_a + f_b = f_c + f_d$. Each of these four-wave mixing processes is sensitive to the phases of the comb modes, which makes it difficult to predict the amplitudes and relative phases of the final comb state. Presently, it is not clear whether the comb phases can be aligned by a power-dependent mode-locking mechanism like in conventional mode-locked laser frequency combs. Recent work shows the generation of solitons in microresonators [13, 14, 18], which would suggest a spontaneous phase-alignment mechanism.

Here, we show the measurement of phases of the modes in microresonator-based optical frequency combs. Most significantly, we find a variety of microcomb states in which the modes undergo discrete phase steps of $\pi$ and $\pi/2$ in different parts of the optical spectrum. The measured comb states are linked to pulses in time domain that are separated by rational fractions of the cavity round-trip time. As a step towards understanding these discrete phase steps, we introduce a new technique to measure the detuning of frequency comb modes from their respective resonator modes in a hot resonator (i.e. while generating optical frequency combs). This measurement utilizes a weak counter-propagating probe laser that does not disturb the comb generation process. We find comb states with all modes "red detuned" (comb mode at lower frequency than microresonator mode) from their resonances. This is a surprising result since the employed microresonators are thermally unstable in the presence of a slightly red-detuned laser, suggesting a nonlinear feedback mechanism on the resonator modes during comb generation.

In addition, the resonator mode detuning measurements provide important input for theoretical simulations of the frequency comb states. The last part of the paper shows these simulations, based on the Lugiato-Lefever equation (a nonlinear, damped Schrödinger equation) [19-22]. Many aspects of the measured comb spectra and phase profiles are in agreement with the simulations.

**Results**

The experimental setup for the measurement of phase and amplitudes of a microresonator-based optical frequency comb is shown in Fig. 1. The employed whispering gallery mode microresonator is a fused-silica microrod [23, 24] with a mode spacing of ~25.6 GHz and a corresponding diameter of around 2.6 mm. The optical Q-factor is $1.5 \times 10^8$. Light is coupled into and out of the resonator using a tapered optical fiber. An optical frequency comb is generated by thermally locking [25] an amplified external cavity diode laser (~100 mW) to a microresonator mode. The high circulating power in the microresonator leads to the generation of an optical frequency comb via four-wave mixing [5, 6], which is coupled out through the same tapered fiber that is used to couple light into the resonator. The generated optical frequency comb is sent through a liquid crystal array-based waveshaper that allows independent control of both the phase and amplitude of each comb mode. In a first step, a computer controlled feedback loop flattens the amplitude of the comb modes using simultaneously measured optical spectra as a reference. In the next step, a nonlinear optical intensity autocorrelator is used to measure the time domain response [26-28, 7, 29] and determine the peak optical power. Another feedback loop iterates the phases of the comb modes until the peak response of the autocorrelator is maximized (corresponding to all comb modes being in phase at a fixed time within each cavity round-trip time). We define the instantaneous field amplitude $A_n$ of the n-th comb mode at the frequency $f_n$ as

$$A_n = A_{n0} \sin(2\pi f_n t + \tilde{\phi}_n) \; ,$$

with the field amplitude $A_{n0}$ and the phase $\tilde{\phi}_n$. The phases can be expanded as

$$\tilde{\phi}_n = \phi_{\text{meas},n} - \phi_{\text{setup},n} = \phi_0 + n\,\phi_1 + \phi_n \; ,$$

with the measured phases $\phi_{\text{meas}}$ and the setup-induced phase shift $\phi_{\text{setup}}$. The linear part $(\phi_0 + n\,\phi_1)$ of the phase just shifts the generated waveform in time (neglecting carrier-envelope effects). Thus, the value of interest in our measurements is $\phi_n$, which we obtain by subtracting setup dispersion $\phi_{\text{setup}}$ and linear slope $(\phi_0 + n\,\phi_1)$ from the measured phases $\phi_{\text{meas}}$. In order to measure the setup dispersion $\phi_{\text{setup}}$, a Fourier-limited reference pulse is sent through the same setup and recompressed with the waveshaper. The retrieved phases $\phi_{\text{setup}}$ that recompress the pulse to its original length directly correspond to the setup dispersion.

The phase optimization works reliably by iterating the phase of one comb mode while maintaining all other phases constant. The iterated phase will be set to a value that increases the peak power of any existing amplitude modulation in the autocorrelation signal. This is done sequentially for all the comb modes. Running the optimization approximately 3 to 4 times for all the comb modes is sufficient to find the global maximum for the autocorrelation response, in which all modes are in phase [30, 29].

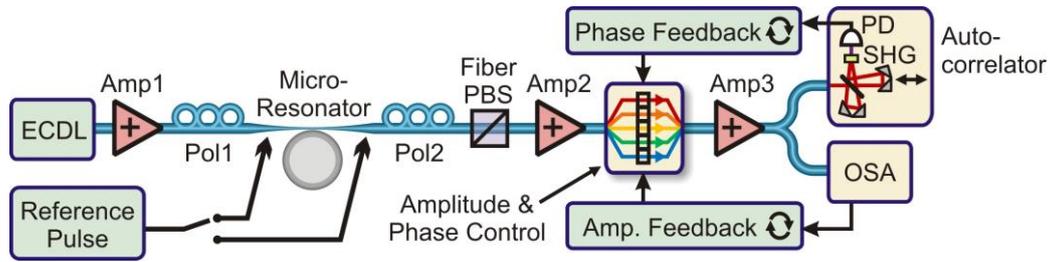

**Figure 1 | Experimental setup for phase measurement of optical frequency comb modes.** An amplified external cavity diode laser generates an optical frequency comb in a microresonator. The comb output is sent into a waveshaper that allows to control phases and amplitudes of individual comb modes. A first feedback loop is used to flatten the amplitude of the comb modes. Subsequently, the phases of the comb modes are retrieved by iteratively changing the phases until the shortest possible pulse with highest peak power is measured in an autocorrelator. The measured phases are corrected for the dispersion of the setup, which is measured with a reference pulse. ECDL = external cavity diode laser, Amp = Erbium doped fiber amplifier, Pol = polarization controller, PBS = polarizing beam splitter, PD = photo diode, SHG = second harmonic generation, OSA = optical spectrum analyzer.

In order to most efficiently use the waveshaper's dynamic amplitude range, it is advantageous to suppress excess pump laser light using a fiber coupled polarizing beam splitter. This pump suppression technique works by slightly misaligning the input polarization (using Pol1 in Fig. 1), such that only part of the pump light couples into the desired mode of the resonator. At the resonator output, the part of the light that did not couple into the resonator (because of mismatched polarization) interferes with the delayed pump light that couples back from the resonator, which in turn leads to an effective polarization change of the pump light. Thus, pump light and comb sidebands are in a different polarization state at the taper output. This polarization difference enables quenching only the pump light using another polarization controller and a fiber coupled polarizer (cf. Fig. 1).

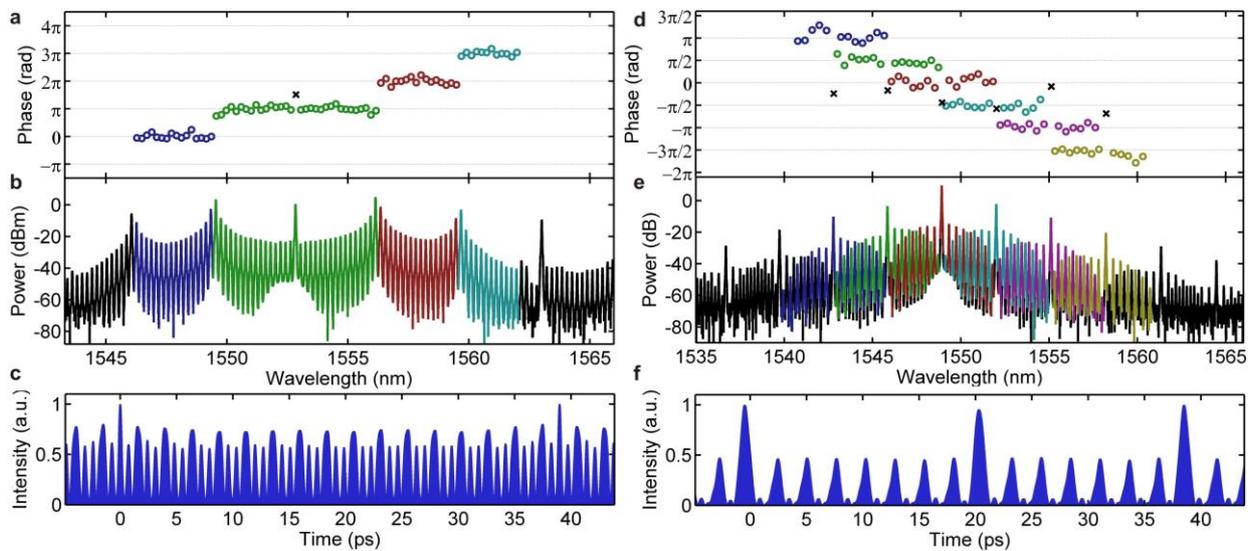

**Figure 2 | "Munich Olympic stadium" comb and interleaved comb state.** Panel a-c and panel d-f show measurements of two different phase locked comb states. The corresponding time domain traces in panel c,f are calculated based on the measured spectral amplitudes and phases. The comb state in panel a-c ("Munich Olympic Stadium"-Comb) shows distinct phase steps of $\pi$ between different sections. The comb state in panel d-f contains interleaved sections with 2xFSR mode spacing and a mutual phase offset of $\pi/2$. The phases of the stronger modes (every 15$^{th}$ FSR, black crosses in the phase plot) are not aligned with the rest of the comb.

Figure 2 shows measurements of two phase locked comb states with ~25.6 GHz mode spacing (~39 ps repetition rate). The data includes the phase of the comb lines, its optical spectrum as well as the time domain representation of the optical signal, calculated from the spectral amplitudes and phases. The comb state in Fig. 2a-c has a characteristic catenary-like spectral envelope and we refer to it as the "Munich Olympic Stadium Comb". The comb initiates from subcomb-bunches [12] that are centered 16 free spectral ranges (FSR) away from the pump laser. Tuning the pump laser more into resonance generates comb modes that fill up the space between bunches until the comb phase-locks similarly to the recently reported self-injection locking phenomenon [30]. The measured phases of the comb modes show distinct steps of $\pi$ between different parts of the spectrum. In the time domain, the comb does not show any strong signs of pulse generation. To the contrary, the different sections of the comb being out of phase by $\pi$ efficiently suppresses any pulse generation. Only this "Munich Olympic Stadium"-comb state grows continuously starting from a bunched comb when tuning the pump laser into resonance.

In contrast, other comb states abruptly emerge from chaotic (not phase locked) comb spectra in a similar way as recently reported in soliton-like mode locking [13, 14]. Although no complicated pump-laser frequency tuning algorithm is required to access them. One of these comb states is shown in Fig. 2d-f and consists of interleaved spectral sections that are offset by $\pi/2$ in phase. Each of the sections is centered around stronger modes (every 15$^{th}$ mode) that have a spacing similar to the spacing of the fundamental combs that are generated before the comb fills in and phase locks when tuning the pump laser into resonance. The calculated time domain trace of this comb state (Fig. 2f) exhibits two stronger pulses per roundtrip time (~39 ps). In addition we observe a faster modulation at 1/15$^{th}$ of the roundtrip time, corresponding to the inverse frequency spacing of the stronger fundamental modes in the spectrum. It is interesting to note that the stronger pulses show up at integer multiples (7x and 8x) of the faster modulation, corresponding to pulse spacings of exactly 7/15$^{th}$ and 8/15$^{th}$ of the intrinsic cavity roundtrip time.

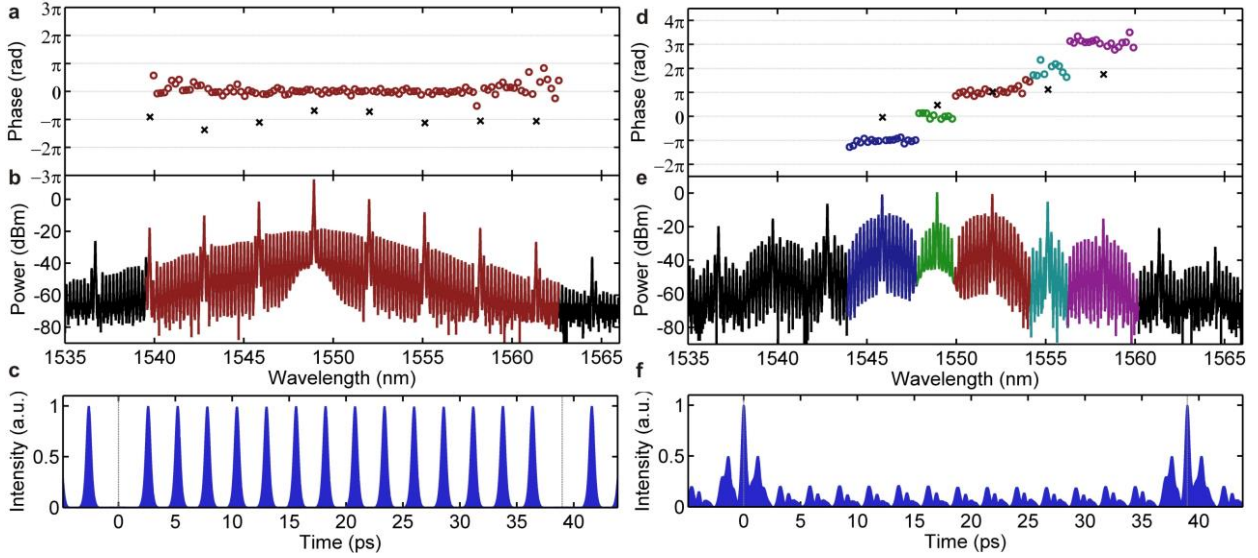

**Figure 3 | Non-interleaved phase-locked comb states.** Panel a-c and panel d-f show experimental data for two more phase locked microresonator comb states. Time domain traces in panel c,f are calculated based on the measured spectral amplitudes and phases. The phases of the comb in panel a-c are aligned, except with the stronger modes (every 15xFSR) having a relative phase of $-\pi \stackrel{\scriptscriptstyle\triangle}{=} \pi$ (black crosses). This leads to a nearly perfect suppression of the expected pulses at $t = 0$ and $t \approx 39$ ps (cavity roundtrip time). The phases of the comb state in panel d-f shows offsets of $+\pi \stackrel{\scriptscriptstyle\triangle}{=} -\pi$ between adjacent sections and has a more complicated temporal structure with triplets of pulses emitted from the microresonator. Spectral amplitudes of this comb state show similarities to an inverted version of the comb in Fig 2b.

Figure 3a-c shows a comb state similar to a previously reported soliton state [13], however, with additional fundamental comb modes at 15xFSR with a phase offset of $-\pi \stackrel{\scriptscriptstyle\triangle}{=} \pi$. The measured comb amplitudes and phases (Fig 3a,b) correspond to a peculiar time domain trace with completely suppressed pulses at the cavity's roundtrip time (0 ps and 39 ps in Fig 3c). The time domain trace only contains pulses at the higher repetition rate corresponding to the inverse frequency spacing of the stronger modes in the spectrum. Calculations show that these suppressed pulses at the cavity's actual roundtrip time require a precise balancing of the powers of the comb modes at 15xFSR in the spectrum with the rest of the comb lines.

One final comb state is shown in Fig 3d-f. The optical spectrum contains different sections of comb modes with a mutual phase offset of $\pi$. To some degree, the spectral envelope of this comb state resembles an inverted version of the "Munich Olympic Stadium" comb shown in Fig 2b. The time domain trace consists of triplets of pulses with a strong central pulse as well as nearly suppressed components at $1/15^{th}$ of the cavity roundtrip time (corresponding to the inverse frequency spacing of the stronger modes in the optical spectrum). Note that all the presented comb states in Fig. 2 and Fig. 3 are generated in the same mode family. Only the coupling position of the tapered fiber is slightly changed in order to generate the different spectra.

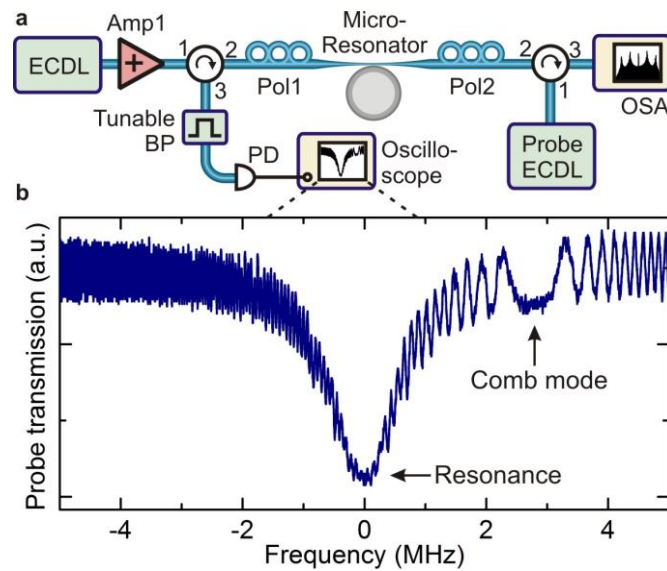

**Figure 4 | Experimental setup for detuning measurements in a hot resonator.** This setup enables the measurement of detunings between optical frequency comb modes and their respective microresonator modes during the comb generation. Panel a. An external cavity diode laser (ECDL) seeds the resonator and generates an optical frequency comb. Simultaneously, a probe ECDL is coupled into the microresonator in backwards direction and swept across one of the modes that is part of the frequency comb. An oscilloscope records the absorption profile of the resonator mode together with the beating between probe laser and the fraction of the light in the comb mode being backreflected in the microresonator (panel b). The time domain axis from the oscilloscope is converted to a frequency axis using the beat frequency as calibration.

Amp = optical amplifier, BP = optical bandpass filter, Pol = polarization controller, OSA = optical spectrum analyzer.

The characteristic spectral amplitudes and phase discontinuities seen in the data of Figs. 2 and 3, along with the phase-locked nature of these states, raise questions about possible anomalies in the resonator mode structure. In order to address this we present measurements of the resonator dispersion and detuning of the optical frequency comb lines from their respective microresonator modes. This measurement contributes important information for numerical modeling of the comb formation process by providing data on the dispersion of the hot resonator, while the comb is generated. The experimental setup for measurement of the comb mode detuning in such a hot resonator is shown in Fig. 4a. The optical frequency comb is generated in the same way as described in Fig. 1 by coupling an amplified external cavity diode laser (ECDL) into a microresonator. However, two optical circulators are added to the setup, which allows the light of an additional low power "probe ECDL" to propagate in backwards direction through the setup. The light of this backwards probe is frequency swept across a microresonator resonance (at a rate of ~20 Hz) and recorded with a fast photodiode. Simultaneously, a small fraction of the comb light that is generated in the same resonance is backreflected (e.g. by small imperfections in the resonator and Rayleigh scattering). This backreflected comb light interferes with the probe laser, generating a beat note that changes in time depending on the current frequency offset between probe laser and comb light. The position of the microcomb mode follows from the position at which the beat note vanishes. An example for such a detuning measurement can be seen in Fig. 4b, showing a Lorentzian-shaped resonance dip combined with the beat note between probe laser and backreflected light from the resonator. This example shows backreflected pump laser light in the far blue detuned regime ("laser frequency" > "resonance frequency") at the beginning of thermal locking [25]. The time axis recorded on the oscilloscope has been replaced with a frequency axis that is directly calibrated by the beat note frequency.

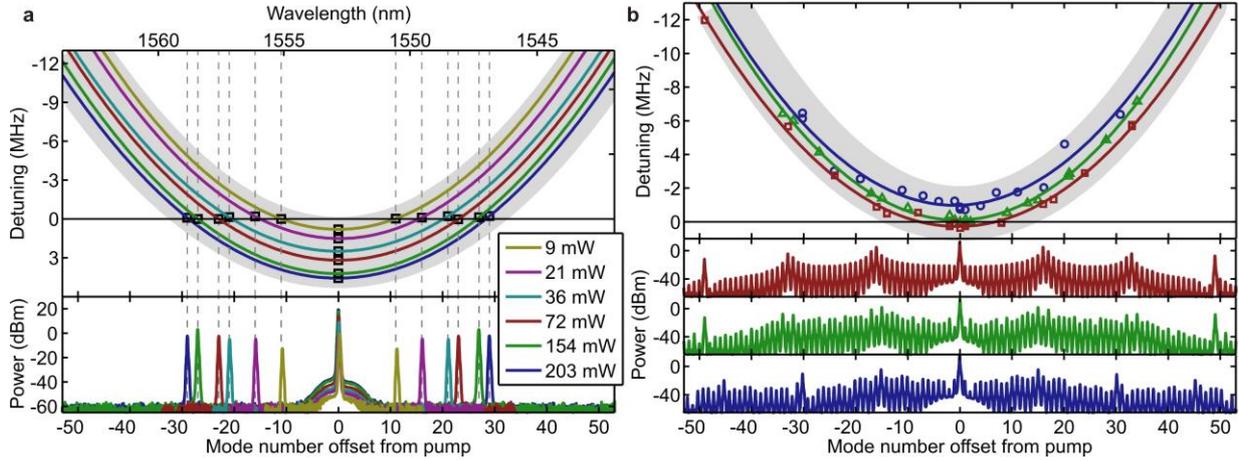

**Figure 5 | Detuning measurements.** Panel a. Measured detunings at the beginning of the comb generation process, with only the first pair of hyper-parametric sidebands being generated. The lower panel shows the optical spectrum at different pump powers. Threshold for comb generation is reached at higher detunings for higher pump powers. The generated comb sidebands are generated close to zero detuning at a spacing determined by the resonator dispersion (broad gray line in the background for comparison). Panel b shows detuning measurements for three different phase-locked comb states in a hot resonator. The curvature of the detuning follows the microresonator dispersion (broad gray line in the background), without obvious shifts due to self- and cross-phase modulation between comb modes. Remarkably, all the comb modes, including the pump laser are red-detuned in the comb states in the two lower spectra (green triangles and blue circles in the detuning plot).

Figure 5 shows detuning measurements for different microresonator states. The data in Fig. 5a are from a state in which the resonator generates only a first pair of parametric sidebands. This is achieved by slowly tuning the pump laser into the resonance from the high frequency side (thermally locking the resonator modes to the laser) until the intracavity power exceeds the threshold for generation of four-wave mixing sidebands. The detuning measurement reveals that the sidebands are generated very close to zero detuning, at multiple FSR away from the pump. Increasing the pump laser power leads to generation of the first four-wave mixing sidebands at larger detunings, with the sidebands still generated close to the center of their respective resonator modes (zero detuning), but further away from the pump laser. A parabolic fit of the detuning versus mode number offset directly reveals the dispersion of the hot resonator. This can be seen by writing the frequency position of the $m^{\text{th}}$ resonator mode as

$$f_m = f_p + m \times f_{rep} - \delta_m \quad ,$$

with the pump frequency $f_p$, the mode number offset with respect to the pump mode $m$, the comb's mode spacing $f_{rep}$, and the detuning of the $m^{\text{th}}$ mode $\delta_m$.

The second order dispersion of the resonator modes can be defined as

$$\widetilde{D}_2 = \frac{d^2 f_m}{dm^2} = -\frac{d^2 \delta_m}{dm^2} \quad ,$$

which corresponds directly to the curvature of the fitted parabolas shown in Fig 5a. In an independent measurement, the cold resonator dispersion (i.e. low power probing of the resonator without comb generation) [31, 12, 32-34, 18] is determined by precisely measuring the change in resonator mode spacing using a probe laser with electro optic modulation sidebands as described in the supplementary information of Ref [30]. The measured cold cavity dispersion of this particular mode family is 11.8 kHz/FSR (anomalous dispersion) and plotted as broad gray line in the background of Fig 5a. The value of the measured hot resonator dispersion obtained from the detuning measurements is (12.2 +/- 0.8) kHz/FSR, which is in good agreement with the cold resonator dispersion. This implies that the cold cavity dispersion of the resonator governs the comb generation process with a negligible contribution of self-phase and cross-phase modulation effects that would shift the position of the modes in the hot resonator during parametric sideband generation [35].

Figure 5b shows detunings and optical spectra for three different phase-locked comb states. The curvature of the detunings is slightly smaller but still in agreement with the cold cavity dispersion (again the gray broad line in the background with a curvature of 11.8 kHz/FSR). The dispersion values from the detunings are (11.6 +/- 1.6) kHz/FSR [blue circles], (11.7 +/- 0.4) kHz/FSR [green triangles] and (10.5 +/- 0.4) kHz/FSR [red squares]. An obvious assumption would be to expect detuning anomalies in the spectral regions that exhibit phase steps.

However, we find that the detuning follows a parabola without obvious deviations, suggesting that the generation of phase-locked comb states is primarily governed by the cold cavity dispersion. More remarkable are the offsets of the detuning parabolas in Fig. 5b. While the pump laser detuning is close to zero in the case of the Munich Olympic Stadium comb (red squares), the offsets are getting smaller (red detuned) for the other two presented comb states with more random spectral envelopes and random phases (cf. Ref. [30]). The pump laser in the comb state in the lowest panel in Fig 5b is significantly red-detuned by 1 MHz, corresponding to 77% of the cavity linewidth (~1.3 MHz). Such a red-detuning is inconsistent with the thermal bistability in a fused silica resonator [25] and indicates the presence of additional unknown optical and/or thermal dynamics that lock the resonator frequencies. A possible explanation is a time dependence of the detuning depending on time domain structure of the circulating light within the resonator [13]. This would allow the microresonator modes to be red detuned, except for times at which the probe light arrives simultaneously with a strong pulse. However, the spectrum in the lowest panel of Fig. 5b is far red-detuned without presence of circulating pulses (cf. autocorrelations of similar comb states in Ref. [30]).

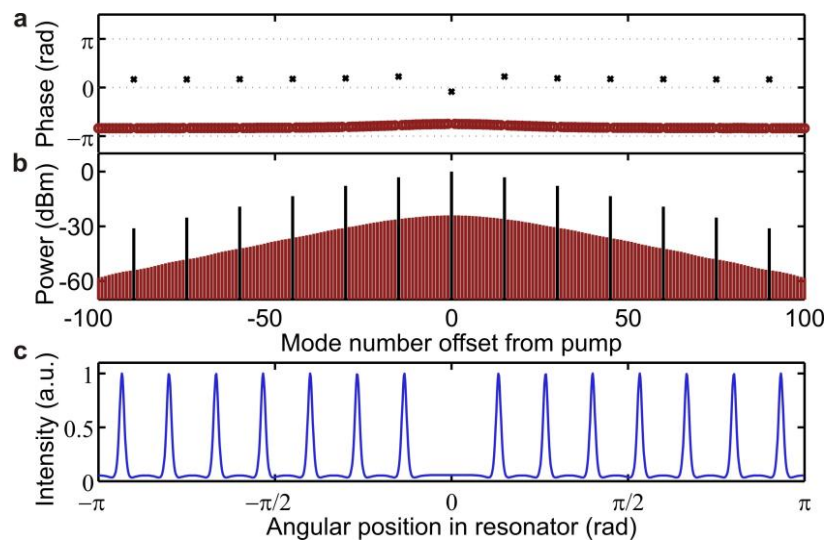

**Figure 6 | Numerical simulation of the comb state in Fig 3a-c.** The figure shows a single solution of the Lugiato-Lefever equation that is in good agreement with the comb state shown in Fig 3a-c. Note that due to the large dispersion value, these solitons will eventually collide in the numerical simulation.

The remaining part of the paper shows numerical simulations of comb states with phase relationships that are similar to those we experimentally measured. The comb simulations are based on the generalized spatiotemporal Lugiato-Lefever equation (LLE) that allows calculating the evolution of the optical field within a resonator [19-22]:

$$\frac{\partial \psi}{\partial \tau} = -(1 + i\alpha)\psi + i|\psi|^2\psi - i\frac{\beta}{2}\frac{\partial^2 \psi}{\partial \theta^2} + F \quad ,$$

where $\psi$ is the normalized field envelope, $\Delta f$ the cavity resonance linewidth, $\tau = \pi \Delta f\, t$ the normalized time, $\alpha = -\frac{2(f_l - f_0)}{\Delta f}$ the normalized cold-cavity detuning between the pump laser frequency $f_l$ and the cold cavity resonance frequency $f_0$, $\beta = -\frac{2\widetilde{D}_2}{\Delta f}$ the dispersion parameter [20], $\theta$ the azimuthal angle along the resonator's circumference, and $F$ the pump laser amplitude normalized to the sideband generation threshold. Because of the Kerr effect, the resonance frequency is shifted, and the cold-cavity detuning $\alpha$ cannot directly be compared to the experimental detuning measurements made at high intra-cavity power. Thus, we introduce the steady state "Kerr" detuning in units of cavity linewidths $\gamma = F^2 - \alpha$. The position of the Kerr-shifted resonance is given by the relation $\alpha = F^2$. A positive $\gamma$ will therefore correspond to a blue detuning and a negative $\gamma$ to a red detuning. In order to interpret the experimental results presented earlier in the paper, numerical simulations of the LLE were performed using a second order split-step Fourier method.

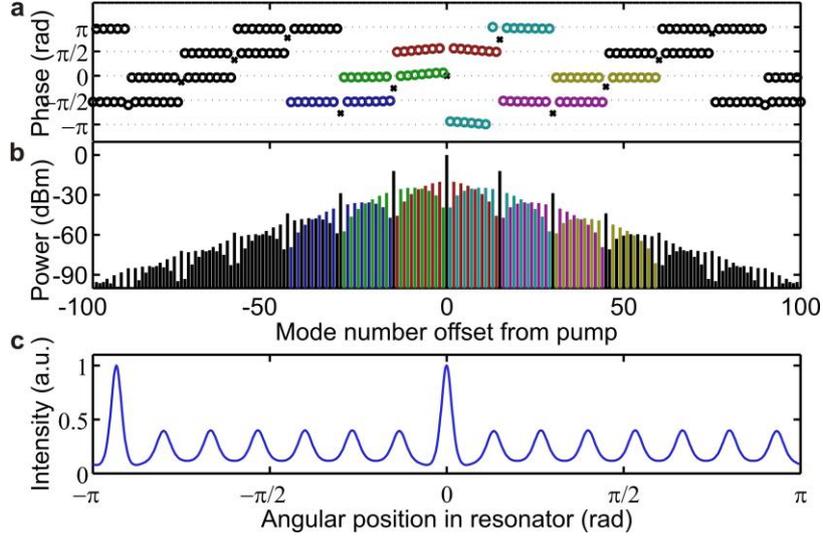

**Figure 7 | Numerical simulation of the interleaved comb state in Fig 2d-f.** The simulation results are a linear combination of two stable solutions of the Lugiato-Lefever equation (with one solution being a comb with lines spaced by 15xFSR, corresponding to the stronger modes in the spectrum).

In the anomalous regime of dispersion, the stationary solutions of the LLE are known to be either solitons or Turing patterns [36, 37], the latter being referred to as "primary combs". In this framework, the experimental comb of Figure 3a-c can be interpreted as a pattern of 14 solitons on a 15-cell grid. Figure 6 shows such a comb simulated in the soliton regime (detuning $\gamma = 0, \alpha = F^2 = 2.5$, and dispersion $\widetilde{D}_2 = 3$ kHz/FSR) with an initial condition consisting of 14 Gaussian pulses positioned along the cavity circumference at angles $\theta_k = 2\pi\, k/15$, with $k = [-7, ..., -1, 1, ..., 7]$. Experimentally, such a comb could be obtained from the initial primary comb corresponding to 15 sub-pulses in the time domain. When the pump laser is tuned into resonance, most of these oscillations turn into solitons, while one of them disappears. While good qualitative agreement can be found between the measurement (Fig. 3a-c) and Fig. 6, it should be noted that this 14-soliton pattern is not stable when using the measured dispersion ($\widetilde{D}_2 = 12$ kHz/FSR). For such a dispersion value, the stable solitons are too long, and packing 14 of them in the cavity makes them interact and collide. This indicates the presence of an additional nonlinear mechanism that is not fully understood.

The experimental results from Figure 2d-f and 3d-f can be thought as the sum of two different solutions of the LLE: a primary comb with 15 FSR-spacing (corresponding to 15 sub-pulses), and a few solitons standing on this pattern. The numerical results of Figure 7 and 8 correspond to the sum of two simulated optical signals using the LLE, using two different values for the detuning parameter: the Turing pattern is obtained for a detuning $\gamma = +2$ (with $\alpha = 0.5$ and $F^2 = 2.5$) which corresponds to a blue detuning in the hot cavity, while the solitons can be generated on both sides of the hot resonance. This latter remark is consistent with our experimental measurement of the detuning presented in Fig 5. To the best of our knowledge, such a linear combination of two solutions of the LLE has only been recently predicted in the case of a dual pumped cavity [38]. In our case, only one pump is used, and the specific mechanism leading to the generation of the observed combs is not fully understood. Further refinements have been proposed to the LLE, by including higher-order dispersion [39], or using the dispersion curve in the supplementary material of Ref. [30] to mimic the effects of mode-crossings. None of these refinements, however, could explain the observed optical spectra. Moreover, in the experiments we do not observe any mode crossings that would disturb the mode family used for comb generation (mode crossings would be also visible in the detuning measurement in Fig. 5b).

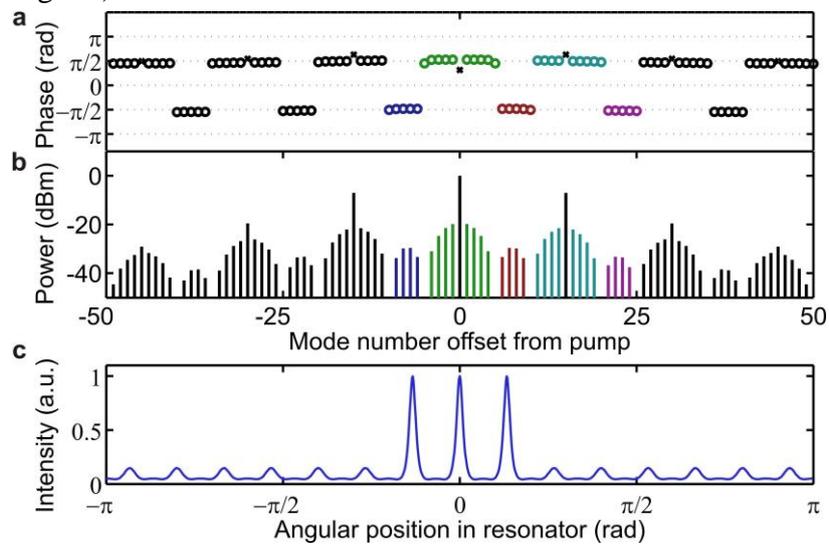

**Figure 8 | Numerical simulation of the comb state in Fig 3d-f.** Linear combination of two stable solutions of the Lugiato-Lefever equation, showing qualitative agreement

with the measured comb state in Fig 3d-f. One of the independent solutions for the simulation of this comb state is a primary comb with modes at every 15$^{th}$ mode (stronger modes in the spectrum in panel b).

**Discussion**

In conclusion we have presented a scheme for measuring optical frequency comb phases, which reveals microresonator comb states with distinct phase steps across the optical spectrum. Moreover, we introduced a method for *in situ* measurements of optical frequency comb mode detunings from the resonator modes in which they are generated. Surprisingly, our measurements show a red-detuning of all the comb modes in some of the phase locked states. The combination of these measurements provides a comprehensive framework for the simulation of microcomb states using the nonlinear Schrödinger equation. We find that the nonlinear dynamics of the comb generation is mostly determined by the dispersion of the resonator, and we show that the LLE is able to qualitatively predict the main features of the complicated spectra and phase profiles of our measurements. The presented detuning measurement technique is promising for future studies in low dispersive resonators, in which it could be the key for experimental evidence on mode-pulling effects resulting from self- and cross-phase modulation between comb modes.

**Acknowledgements:** This work is supported by NIST, the DARPA QuASAR program, the AFOSR and NASA. PD thanks the Humboldt Foundation for support. This paper is a contribution of NIST and is not subject to copyright in the United States.

$^{*}$pascal.delhaye@nist.gov